\documentclass[10pt]{article}
\usepackage{amsmath}
\usepackage{amsfonts}

\usepackage{amsmath,amssymb}
\usepackage{graphicx}
\usepackage{color}
\usepackage{url}

\newtheorem{theorem}{Theorem}
\newtheorem{lemma}{Lemma}
\newtheorem{corollary}{Corollary}

\newtheorem{remark}{Remark}
\newtheorem{conjecture}{Conjecture}

\newcommand{\F}{\ensuremath{\mathbb F}}
\newcommand{\Z}{\ensuremath{\mathbb Z}}

\newcommand{\done}{\hfill $\Box$ }

\newcommand{\Tr}{{{\rm Tr}}}

\newcommand{\ls}[1]
    {\dimen0=\fontdimen6\the\font\lineskip=#1\dimen0
     \advance\lineskip.5\fontdimen5\the\font
     \advance\lineskip-\dimen0
     \lineskiplimit=0.9\lineskip
     \baselineskip=\lineskip
     \advance\baselineskip\dimen0
     \normallineskip\lineskip\normallineskiplimit\lineskiplimit
     \normalbaselineskip\baselineskip
     \ignorespaces}

\begin{document}

\bibliographystyle{abbrv}


\title{Several Classes of Negabent Functions over Finite Fields}
\author{Gaofei Wu
\thanks{G. Wu and X.  Liu are  with the State Key Laboratory of Integrated Service Networks,
Xidian University, Xi'an, 710071, China.
 Email: wugf@nipc.org.cn, liuxf@nipc.org.cn},
Nian Li
\thanks{N. Li is  with the  Department of Informatics, University of Bergen,
 N-5020 Bergen, Norway.
 Email:
nianli.2010@gmail.com.},
Yuqing Zhang
\thanks{Y. Zhang is with the  National Computer Network Intrusion Protection Center, UCAS, Beijing 100043, China.
 Email:
zhangyq@ucas.ac.cn.},
and
Xuefeng Liu}
\date{}
\maketitle
\ls{1.5}

\thispagestyle{plain} \setcounter{page}{1}

\begin{abstract}
Negabent functions as a class of
 generalized bent functions have attracted a lot of
attention recently due to their applications in cryptography and coding theory.
In this paper, we consider the constructions of  negabent functions over finite fields.
First, by using the  compositional inverses of
   certain binomial and trinomial permutations,
  we present several classes of negabent functions of the form
  $f(x)=\Tr_1^n(\lambda x^{2^k+1})+\Tr_1^n(ux)\Tr_1^n(vx)$, where
    $\lambda\in \F_{2^n}$,  $2\leq k\leq n-1$,  $(u,v)\in \F^*_{2^n}\times \F^*_{2^n}$,
and    $\Tr_1^n(\cdot)$ is the   trace function from  $\F_{2^n}$ to $\F_{2}$.
Second, by using Kloosterman sum,  we prove that the condition for
 the cubic monomials given by Zhou and Qu (Cryptogr. Commun., to appear,
DOI 10.1007/s12095-015-0167-0.)
 to be negabent is also necessary.
In addition,   a conjecture on negabent monomials whose  exponents are
  of Niho type is given.

\end{abstract}

\section{Introduction}
 Bent  functions are an important class of Boolean functions
which  were introduced by Rothaus \cite{Rothaus}.
 A Boolean function is called
\emph{bent} if and only if  it has a flat spectrum  with respect to the Walsh-Hadamard transform.
Bent functions have attracted a lot of attention
 due to their applications in
coding theory and cryptography.
As a logical extension of  bent functions,
 Kumar, Scholtz, and Welch \cite{1985} gave the definition of  $p$-ary bent functions from
$\Z_{p}^n$ to $\Z_p$, where $p$ is an integer.
Schmidt \cite{schmidt09} introduced the generalized Boolean
bent functions from $\Z_2^m$ to $\Z_p$ from the viewpoint of cyclic codes over Galois ring.

Motivated by a choice of local unitary transforms that are central to the
structural analysis of pure $n$-qubit stabilizer quantum states,
Riera and Parker \cite{matthew06}  introduced some generalized bent criteria for Boolean functions.
They considered Boolean functions that have a flat spectrum with respect to one or more matrix transforms
 from the $\{I,H,N\}^n$ set of matrices or subsets thereof, where
\begin{footnotesize} $I=  \left ( \begin{array}{rr} 1 & 0 \\ 0 & 1 \end{array} \right )$,
$H= \frac{1}{\sqrt{2}} \left ( \begin{array}{rr} 1 & 1 \\ 1 & -1 \end{array} \right )$,\end{footnotesize} and
\begin{footnotesize}$N= \frac{1}{\sqrt{2}} \left ( \begin{array}{rr} 1 & \sqrt{-1} \\ 1 & -\sqrt{-1} \end{array} \right )$.\end{footnotesize}
 A $2^n \times 2^n$ transform matrix, $U$, is in the set $\{I,H,N\}^n$ if it can be
written as $U = U_0 \otimes U_1 \otimes \ldots \otimes U_{n-1} = \bigotimes_{j=0}^{n-1} U_j$,
where $U_j \in \{I,H,N\}$ and $\otimes$ is the tensor product. Thus $\{I,H,N\}^n$ is
 a set of $3^n$ transform matrices.
A   {\em negabent}  function is a Boolean function which has flat spectrum with respect
to the  {\em negaHadamard}, $N^{\otimes n}$, transform.
{\em Bent-negabent}  functions are 
Boolean  functions  that  are both bent and negabent.
In 2007, Parker and Pott \cite{matthew07} gave an important connection between
bent and negabent functions, and showed that if $n$ is even, then one can
obtain  negabent functions from any bent ones. By using this connection,
St${\rm \check{a}}$nic${\rm \check{a}}$ \cite{gang11} gave a class of $n$-variable
 bent-negabent functions
with algebraic degree $\frac{n}{4}+1$.
 Su, Pott, and Tang \cite {su13} considered the negaHadamard spectra of negabent functions, and
constructed a class of bent-negabent functions with optimal algebraic degree by using complete
permutation polynomials.  Recently, Zhang, Wei, and Pasalic \cite{zhang15}
used the indirect sum construction
 proposed by Carlet \cite{carlet04} to
 construct  the first
class of bent-negabent functions which are not in the completed Maiorana-McFarland class.
On the other hand, it is also important to construct negabent functions over finite fields.
 Sarkar \cite{sarkar12} considered
  negabent functions over finite fields, and characterized all the quadratic negabent monomials over finite fields.
  Recently,
 Zhou and Qu \cite{zhou15} gave a class of cubic monomial negabent functions and a class of
 cubic negabent polynomials over finite fields.

 In this paper,
  we first give the necessary and sufficient conditions for the functions
  $\Tr_1^k(\lambda x^{2^k+1})+\Tr_1^n(ux)\Tr_1^n(vx)$ to be negabent, where
   $n=2k$, $\lambda\in \F_{2^k}$,  and $(u,v)\in \F^*_{2^n}\times \F^*_{2^n}$.
   Then by using some permutation trinomials over $\F_{2^n}$, we present
    some classes of
   negabent functions of the form  $\Tr_1^n(\lambda x^{2^k+1})+\Tr_1^n(ux)\Tr_1^n(vx)$,  where
   $0<k<n$.
 Third, we show that the condition for the cubic monomials given by Zhou and Qu \cite{zhou15}
 to be negabent is also necessary. Kloosterman sum plays  an important role in the proof.
 In addition, we present  a conjecture on negabent monomials whose  exponents are of Niho type.

The remainder  of this paper is organized as follows.   In
Section \ref{secpre}, some preliminaries including Kloosterman sum and permutation polynomials
over finite fields  are introduced.  In Section
\ref{secnegapoly}, by using the compositional inverses of some  binomial and trinomial permutations,
  several classes of negabent functions of the form
 $\Tr_1^n(\lambda x^{2^k+1})+\Tr_1^n(ux)\Tr_1^n(vx)$ are given.
 A class of negabent monomials over finite fields is
  considered in Section \ref{secnegamono},
   and some concluding remarks are given in Section \ref{seccon}.



\section{Preliminaries}\label{secpre}

A Boolean function $f(x)$ is a mapping from $\F_2^n$ to $\F_2$. The Walsh-Hadamard transform of a function
$f(x)$ at $a\in\F_2^n $ is defined by
 $$
W_{f}(a)=\sum_{x\in{\mathbb F}_{2}^n}(-1)^{f(x)+a\cdot x},
$$
where $a\cdot x$ is the standard inner product. If for any  $a\in\F_2^n $, $|W_{f}(a)| =2^{\frac{n}{2}}$, then $f(x)$ is called a \emph{bent} function.
It is known that
an $n$-variable Boolean  function   $f(x)$ is bent if and only if $f(x)+f(x+a)$ is balanced for all nonzero $a\in \F_2^n$.
In \cite{matthew06}, Riera and Parker introduced the notion of  negabent function.  The negaHadamard transform
 of $f(x)$ at $a\in \F_2^n$ is defined by
$$
N_f(a)=\sum_{x\in\F_2^n}(-1)^{f(x)+a\cdot x}\sqrt{-1}^{wt(x)},
$$
where $wt(x)$ is the weight of the vector $x=(x_0,x_1,\cdots,x_{n-1})$, i.e., $wt(x)=\#\{i~|~x_i=1, i\in \Z_n\}.$
A function  $f(x)$ is called a  \emph{negabent} function if
$|N_f(a)|=2^{\frac{n}{2}}$ for all $a\in \F_2^n$.
Similarly, a function $f(x)$ is negabent if and only if $f(x)+f(x+a)+a\cdot x$ is balanced
for all nonzero $a\in \F_2^n$.

In this paper, we focus on negabent functions over finite fields.  It is well known that
the vector space $\F_{2}^n$ is homomorphic to the finite field $\F_{2^n}. $  Let $k$ be an integer
such that $k|n$. The trace function from
${\mathbb F}_{2^n} $ onto ${\mathbb F}_{2^k}$ is defined by
$$
\Tr_k^n(x)=\sum_{i=0}^{n/k-1}x^{2^{ik}}, \, x\in {\mathbb F}_{2^n}.
$$
If $k=1$, we call $\Tr_1^n(x)$  the absolute trace function from
${\mathbb F}_{2^n} $ to ${\mathbb F}_{2}$. Let $\{\alpha_1,\alpha_2,\cdots, \alpha_n\}$  be  a self dual basis of
$\F_{2^n}$ over $\F_2. $ Let $x=\sum\limits_{i=1}^n x_i\alpha_i$ and $a=\sum\limits_{i=1}^n a_i\alpha_i$, then
$\Tr_1^n(ax)=\sum\limits_{i=1}^n a_ix_i=a\cdot x$. Thus  we have the following equivalent  definition of negabent
functions  over finite fields, which was first introduced by Sarkar in \cite{sarkar12}.

%


\begin{theorem}\cite{sarkar12}\label{negath}
Let $f(x)$ be a Boolean function from $\F_{2^n}$ to $\F_2$.
Then $f(x)$ is negabent if and only if
$$
\sum_{x\in \F_{2^n}}(-1)^{f(x)+f(x+a)+\Tr_1^n(ax)}=0
$$
for all nonzero a in $\F_{2^n}$.
\end{theorem}


In what follows we present some results on certain exponential sums and permutation polynomials
 over finite fields,  which will play an important  role in our proofs.
%
%
%

%


Let $a,b\in \F_{2^n}$, the Kloosterman sum over $\F_{2^n}$ is defined by
\[K_n(a,b)=\sum\limits_{x\in \F_{2^n}^*}(-1)^{\Tr_1^n(ax+bx^{-1})}.\]

%

\begin{lemma}\label{kloobound}\cite[Theorem 5.45]{Lidl97}
If $a,b\in\F_{2^n} $ are not both zero,  then the Kloosterman sum satisfies
$$
|K_n(a,b)|\leq 2 \sqrt{{2^n}}.
$$
\end{lemma}

\begin{lemma}\label{lemma-kloo}
Let $k$ be a positive integer and $q=2^k$.  For any $b\in \F_q^*$ and $c\in \F_q^*$,  define $A=\#\{x\in \F_q^*\,|\,\Tr_1^k(bx)=0, \Tr_1^k(cx^{-1})=1\}$.
Then $A>0$ if $k>2$.
\end{lemma}

{\em Proof:} Let
 $B=\#\{x\in \F_q^*\,|\,\Tr_1^k(bx)=1, \Tr_1^k(cx^{-1})=0\}$,
  $C=\#\{x\in \F_q^*\,|\,\Tr_1^k(bx)=0, \Tr_1^k(cx^{-1})=0\}$, and
   $D=\#\{x\in \F_q^*\,|\,\Tr_1^k(bx)=1, \Tr_1^k(cx^{-1})=1\}$.
   Then it is readily to verify that  $A+C=2^{k-1}-1, B+D=2^{k-1}$ and $A+D=2^{k-1}$. This together with Lemma \ref{kloobound}, i.e., $|A+B-C-D|\leq 2\sqrt{q}$, leads to $|4A-2^k+1|\leq 2\sqrt{q}$, which implies that $A>0$ if $k>2$. This completes the proof.
\done

A polynomial $f\in {\mathbb F}_{q}[x]$ is called a permutation polynomial  if
the associated polynomial mapping
$f:c\mapsto f(c)$ from ${\mathbb F}_{q}$ to itself is a permutation of
${\mathbb F}_{q}$ \cite{Lidl97}.

\begin{lemma}\label{ppcriteria}\cite[p.118]{Lidl97}
Let $q$  be a prime power and $f(x)=\sum_{i=0}^{m-1}a_ix^{q^i}\in\F_q[x]$.
Then $f(x)$ is a permutation polynomial over $\F_{q^m}$ if and only if
$\gcd(\sum_{i=0}^{m-1}a_ix^i, x^m-1)=1$. Moreover, if $g(x)$ is the compositional inverse of
$f(x)$, i.e., $f(g(x)) \equiv x\mod (x^{q^m}-x)$, then $g(x)$ is a $q$-polynomial over $\F_q$.
\end{lemma}

\begin{lemma}\label{permu1}
Let $k$ be a positive integer and $f(x)=x+x^{2^k}+x^{2^{2k}}$, then $f(x)$ is a permutation polynomial over $\F_{2^{n}}$ if and only if $\gcd(n,3k)=\gcd(n,k)$. Further, let $g(x)$ be the compositional  inverse of $f(x)$. Then $g(x)$ is a $2$-polynomial over $\F_2$ and $\Tr_1^n(g(x))=\Tr_1^n(x)$.
\end{lemma}
{\em Proof:}
According to Lemma \ref{ppcriteria}, $f(x)$ is a permutation polynomial over $\F_{2^{n}}$ if and only if $\gcd(\frac{x^{3k}-1}{x^k-1}, x^n-1)=1$.
Note that $\gcd( \frac{x^{3k}-1}{x^k-1}, x^k-1 )=\gcd(3, x^k-1)=1$. This implies that $\gcd(x^{3k}-1, x^n-1)=\gcd(\frac{x^{3k}-1}{x^k-1}, x^n-1)\cdot \gcd(x^{k}-1, x^n-1)$ which leads to
$\gcd(\frac{x^{3k}-1}{x^k-1}, x^n-1)=\frac{x^{\gcd(n,3k)}-1}{x^{\gcd(n,k)}-1}$.
Thus, $f(x)$ is a permutation polynomial over $\F_{2^{n}}$ if and only if $\gcd(n,3k)=\gcd(n,k)$.

If $g(x)$ is the compositional inverse of $f(x)$, then we have $g(x)$ is a $2$-polynomial over $\F_2$ due to Lemma \ref{ppcriteria}. Moreover, we have $g(1)=1$ since $f(1)=1$, i.e., $g(x)$ has odd number of terms. This leads to $\Tr_1^n(g(x))=\Tr_1^n(x)$ since $g(x)$ is a $2$-polynomial over $\F_2$. This completes the proof.
\done

\begin{lemma}\label{permu2}
Let $n=rk$ and  $f(x)=\lambda x+x^{2^k}+\lambda x^{2^{2k}}$, where $r, k$ are positive integers and $\lambda\in\F_{2^k}^*$.
Then $f(x)$  is a permutation polynomial over $\F_{2^{n}}$ if and only if $\gcd(\lambda+x+\lambda x^{2},x^r-1)=1$.  Further, let $g(x)$ be
the compositional  inverse of $f(x)$. Then $g(x)$ is a $2^k$-polynomial over $\F_{2^k}$ and $\Tr_1^n(g(x))=\Tr_1^n(x)$.
\end{lemma}
{\em Proof:}
Note that $f(x)$ is a $2^k$-polynomial over $\F_{2^k}$. Thus the first assert follows directly from Lemma \ref{ppcriteria}. Further, by Lemma \ref{ppcriteria} we have that $g(x)$ is also a $2^k$-polynomial over $\F_{2^k}$ if $g(x)$ is the compositional inverse of $f(x)$.
Suppose that $g(x)=\sum_{i=0}^{r-1}c_ix^{2^{ki}}$, where $c_i\in\F_{2^k}$. Then, we have
$\Tr_1^n(g(x))=\Tr_1^k(\Tr_k^{rk}(g(x)))=\Tr_1^k(\Tr_k^{rk}(\sum_{i=0}^{r-1}c_ix^{2^{ki}}))=
 \Tr_1^k(\sum_{i=0}^{r-1}c_i\Tr_k^{rk}(x^{2^{ki}} ))=
 \Tr_1^k(g(1)\Tr_k^{rk}(x))$. Then the result follows from the fact that $g(1)=1$ since $f(1)=1$. This completes the proof.
\done

%
%
%
%

\section{Some  classes of negabent polynomials}\label{secnegapoly}


In this section,
by using some permutation polynomials over
$\F_{2^n}$,   we present several classes of negabent functions of the form $ \Tr_1^n(\lambda x^{2^k+1})+\Tr_1^n(ux)\Tr_1^n(vx)$
over   $\F_{2^n}$,  where $2 \leq k \leq n-1$,  $\lambda\in \F_{2^n}$,  and $(u,v)\in \F^*_{2^n}\times \F^*_{2^n}$.

\begin{theorem}\label{negapoly1}
Let $n=2k$, $\lambda\in \F_{2^k}$ and $(u,v)\in \F^*_{2^n}\times \F^*_{2^n}$.
 Then $f(x)=\Tr_1^k(\lambda x^{2^k+1})+\Tr_1^n(ux)\Tr_1^n(vx)$
is negabent on $\F_{2^n}$ if and only if one of the following  conditions is
satisfied:
\begin{enumerate}
 \item $\lambda\neq 1$, $(\Tr_1^n(\frac{u}{1+\lambda}),\Tr_1^n(\frac{(\lambda u^{2^k}+u)v}{1+\lambda^2}),\Tr_1^n(\frac{v}{1+\lambda}))\in\{(0,0,0), (0,0,1), (1,0,0), (1,1,1)\}$;
  \item $\lambda= 1, \, k=2,\, u, v, u+v\not\in\F_{2^k}$;
  \item  $\lambda=1, k=1$, $u\ne v$.
\end{enumerate}
\end{theorem}
{\em Proof:}
According to Theorem \ref{negath}, to complete this proof, it is sufficient to prove that $f(x)+f(x+a)+\Tr_1^n(ax)$ is balanced for all nonzero $a\in\F_{2^n}$ if and only if $\lambda, u, v$ satisfy one of the conditions given in Theorem \ref{negapoly1}.
A direct calculation gives
 \begin{eqnarray*}
 f(x)+f(x+a)+\Tr_1^n(ax) &=&\Tr_1^k(\lambda(a^{2^k}x+ax^{2^k}))+\Tr_1^n(ua)\Tr_1^n(vx)+\Tr_1^n(va)\Tr_1^n(ux)+\Tr_1^n(ax)\\
 &&+\Tr_1^k(\lambda a^{2^k+1})
 +\Tr_1^n(ua)\Tr_1^n(va)\\
 &=&\Tr_1^n((\lambda a^{2^k}+a)x)+\Tr_1^n(v\Tr_1^n(ua)x)+\Tr_1^n(u\Tr_1^n(va)x)\\
 &&+\Tr_1^k(\lambda a^{2^k+1})
 +\Tr_1^n(ua)\Tr_1^n(va).
 \end{eqnarray*}
This implies that $f(x)+f(x+a)+\Tr_1^n(ax)$ is balanced if and only if $\lambda a^{2^k}+a+v\Tr_1^n(ua)+u\Tr_1^n(va)\ne 0$.
Notice that $\lambda a^{2^k}+a$ is a $2^k$-polynomial and $\gcd(\lambda a^k+1, a^{2k}+1)=\gcd(\lambda a^k+1, (a^{k}+1)^2)=\gcd(\lambda+1, a^k+1)=1$ only if $\lambda\not=1$. This together with Lemma \ref{ppcriteria} shows that $\lambda a^{2^k}+a$ is permutation polynomial if $\lambda\not=1$. Moreover, for any $\lambda\not=1$ and $b\in\F_{2^n}$, if $\lambda a^{2^k}+a=b$, then one gets $\lambda a+a^{2^k}=b^{2^k}$ since $n=2k$ and $\lambda\in \F_{2^k}$. These two identities lead to
\begin{eqnarray}\label{eq-inverse-a}
  a=\frac{b+\lambda b^{2^k}}{\lambda^2 +1},
\end{eqnarray}
which is the unique solution to $\lambda a^{2^k}+a=b$.

For simplicity, define $h(a)=\lambda a^{2^k}+a+v\Tr_1^n(ua)+u\Tr_1^n(va)$. Then by \eqref{eq-inverse-a}, for $\lambda\not=1$ we have
\begin{enumerate}
  \item [1)] $(\Tr_1^n(ua), \Tr_1^n(va))=(0,0)$: For this case, $h(a)=0$ has the only solution $a=0$.
  \item [2)] $(\Tr_1^n(ua), \Tr_1^n(va))=(0,1)$: By \eqref{eq-inverse-a}, $a=\frac{u+\lambda u^{2^k}}{\lambda^2 +1}$ is the unique solution to $\lambda a^{2^k}+a+u=0$. Note that $\Tr_1^n(ua)=\Tr_1^n(u\cdot\frac{\lambda u^{2^k}+u}{1+\lambda^2})=\Tr_1^n(\frac{\lambda u^{2^k+1}}{1+\lambda^2})+\Tr_1^n(\frac{u^2}{1+\lambda^2})=\Tr_1^n(\frac{u}{1+\lambda})$ since $n=2k$ and $\frac{\lambda u^{2^k+1}}{1+\lambda^2}\in\F_{2^k}$. Thus, in this case $h(a)=0$ has the only solution $a=\frac{u+\lambda u^{2^k}}{\lambda^2 +1}$ if and only if $\Tr_1^n(\frac{u}{1+\lambda})=0$ and $\Tr_1^n(va)=\Tr_1^n(v\cdot \frac{\lambda u^{2^k}+u}{1+\lambda^2})=1$.
  \item [3)] $(\Tr_1^n(ua), \Tr_1^n(va))=(1,0)$: Similar as above, for this case $h(a)=0$ has the only solution $a=\frac{v+\lambda v^{2^k}}{\lambda^2 +1}$ if and only if $\Tr_1^n(\frac{v}{1+\lambda})=0$ and $\Tr_1^n(ua)=\Tr_1^n(u\cdot\frac{\lambda v^{2^k}+v}{1+\lambda^2})=1$.
  \item [4)] $(\Tr_1^n(ua), \Tr_1^n(va))=(1,1)$: In this case, $a=\frac{u+v+\lambda (u+v)^{2^k}}{\lambda^2 +1}$ is the unique solution to $\lambda a^{2^k}+a+u+v=0$ due to \eqref{eq-inverse-a}. By the same techniques used in Cases 2) and 3) one can conclude that $h(a)=0$ has the only solution if and only if $\Tr_1^n(\frac{u}{1+\lambda}+\frac{(\lambda v^{2^k}+v)u}{1+\lambda^2})=1$ and $\Tr_1^n(\frac{v}{1+\lambda}+\frac{(\lambda u^{2^k}+u)v}{1+\lambda^2})=1$.
\end{enumerate}
Notice that $\Tr_1^n(\frac{(\lambda v^{2^k}+v)u}{1+\lambda^2})=\Tr_1^n(\frac{(\lambda vu^{2^k})^{2^k}}{(1+\lambda^2)^{2^k}})+\Tr_1^n(\frac{vu}{1+\lambda^2})=\Tr_1^n(\frac{\lambda vu^{2^k}}{1+\lambda^2})+\Tr_1^n(\frac{vu}{1+\lambda^2})=\Tr_1^n(\frac{(\lambda u^{2^k}+u)v}{1+\lambda^2})$ due to $n=2k$ and $\lambda\in\F_{2^k}$. Therefore, if $\lambda\ne 1$, by combining Cases 1)--4), one has that $h(a)=\lambda a^{2^k}+a+v\Tr_1^n(ua)+u\Tr_1^n(va)\ne 0$ for any nonzero $a\in\F_{2^n}$ if and only if the first condition in Theorem \ref{negapoly1} is satisfied.

Now we consider the case of $\lambda=1$. First we discuss the number of solutions of $h(a)=\lambda a^{2^k}+a+v\Tr_1^n(ua)+u\Tr_1^n(va)$ under the condition $(\Tr_1^n(ua), \Tr_1^n(va))=(0,0)$. In this case, $h(a)=0$ is equivalent to $a\in\F_{2^k}$. Let $N(u,v)$ denote the number of nonzero $a\in\F_{2^k}$ such that $(\Tr_1^n(ua), \Tr_1^n(va))=(0,0)$, where $(u,v)\in \F^*_{2^n}\times \F^*_{2^n}$. Then, according to the balanced property of the trace function and the fact that $(\Tr_1^n(ua), \Tr_1^n(va))=(\Tr_1^k(a(u+u^{2^k})), \Tr_1^k(a(v+v^{2^k})))$, it can be readily verified that $N(u,v)=2^k-1$ if $u,v\in\F_{2^k}$, $N(u,v)=2^{k-1}-1$ if exactly one of $u,v$ belongs to $\F_{2^k}$, $N(u,v)=2^{k-1}-1$ if $u,v\not\in\F_{2^k}$ with $u+v\in\F_{2^k}$ and $N(u,v)=2^{k-2}-1$ if $u,v, u+v\not\in\F_{2^k}$ respectively. This implies that $h(a)=0$ under the condition $(\Tr_1^n(ua), \Tr_1^n(va))=(0,0)$ has at least one nonzero solution for any given $u,v\in\F_{2^n}$ if $k>2$, i.e., $f(x)$ cannot be negabent if $\lambda=1$ and $k>2$. The conditions on $u,v\in \F_{2^n}$ such that $f(x)$ is negabent for $k=1,2$ can be easily verified based on a simple discussion. This completes the proof. \done

\begin{remark}
Let $u=v$ in Theorem \ref{negapoly1}, then $f(x)$ is negabent on $\F_{2^n}$ if and only if
$\lambda\neq 1$, which is  Proposition 5 in \cite{sarkar14}.
\end{remark}

\begin{corollary} Let $f(x)$ with $u\neq v$ be given as in Theorem \ref{negapoly1} and $\mathbb{N}_{\lambda}$ denote the number of ordered  pairs $(u,v)$ such that $f(x)$ is negabent. Then $\mathbb{N}_{\lambda}=(2^{n-1}-2)(2^n-1)$ for any fixed $\lambda\neq 1$ and $\mathbb{N}_1=6, 96$ for $k=1, 2$ respectively.
\end{corollary}
{\em Proof: } We only give the proof for $\lambda \neq1$ since the proof for $\lambda=1$ is trivial due to Theorem \ref{negapoly1}. For $\lambda \neq1$, we first determine the number of ordered  pairs $(u,v)$ such that
$(\Tr_1^n(\frac{u}{1+\lambda}),\Tr_1^n(\frac{(\lambda u^{2^k}+u)v}{1+\lambda^2}),\Tr_1^n(\frac{v}{1+\lambda}))\in\{(0,0,0), (0,0,1)\}$.
 Note that $(\Tr_1^n(\frac{u}{1+\lambda}),\Tr_1^n(\frac{(\lambda u^{2^k}+u)v}{1+\lambda^2}),\Tr_1^n(\frac{v}{1+\lambda}))\in\{(0,0,0), (0,0,1)\}$ is equivalent to $(\Tr_1^n(\frac{u}{1+\lambda}),\Tr_1^n(\frac{(\lambda u^{2^k}+u)v}{1+\lambda^2}))=(0, 0)$. Clearly, the number of $u\in\F_{2^n}^*$ satisfying $\Tr_1^n(\frac{u}{1+\lambda})=0 $ is $2^{n-1}-1$, and for each such $u$, there are $2^{n-1}-2$  $v$'s in $\F_{2^n}^*\setminus \{u\}$  such that $\Tr_1^n(\frac{(\lambda u^{2^k}+u)v}{1+\lambda^2})=0$. Thus, in this case we get $(2^{n-1}-1)(2^{n-1}-2)$ ordered  pairs $(u,v)$ such that $f(x)$ is negabent.

Next we count the number of the pairs $(u,v)$ such that $(\Tr_1^n(\frac{u}{1+\lambda}),\Tr_1^n(\frac{(\lambda u^{2^k}+u)v}{1+\lambda^2}),\Tr_1^n(\frac{v}{1+\lambda}))\in\{(1,0,0), (1,1,1)\}$,
which is equivalent to counting the number of the pairs $(u,v)$ satisfying $\Tr_1^n(\frac{u}{1+\lambda})=1$ and $\Tr_1^n(\frac{(\lambda u^{2^k}+u)v}{1+\lambda^2})+
\Tr_1^n(\frac{v}{1+\lambda})=\Tr_1^n(\frac{(\lambda u^{2^k}+u+1+\lambda)v}{1+\lambda^2})=0$.
Similar as above, for this case the number of $u\in\F_{2^n}^*$ satisfying $\Tr_1^n(\frac{u}{1+\lambda})=1 $ is $2^{n-1}$, and for each such $u$, there are $2^{n-1}-2$  $v$'s in $\F_{2^n}^*\setminus \{u\}$  such that $\Tr_1^n(\frac{(\lambda u^{2^k}+u+1+\lambda)v}{1+\lambda^2})=0$, i.e., we have $2^{n-1}(2^{n-1}-2)$ ordered  pairs $(u,v)$ such that $f(x)$ is negabent. This completes the proof.
%
%
\done

The function $f(x)$ in Theorem \ref{negapoly1} has been investigated recently by Mesnager \cite{Mesnager14} in order to construct new classes of bent functions.

\begin{theorem}\label{bentpoly}\cite{Mesnager14}
Let $n=2k$, $\lambda\in\F_{2^k}^*$ and $(u,v)\in \F_{2^n}^*\times\F_{2^n}^*$, then
 $f(x)=\Tr_1^k(\lambda x^{2^k+1})+\Tr_1^n(ux)\Tr_1^n(vx)$ is bent if and only if
 $ \Tr_1^n(\lambda^{-1}u^{2^k}v)=0$.
\end{theorem}

Combining Theorem \ref{negapoly1} and Theorem \ref{bentpoly}, we have the following corollary.

\begin{corollary}
Let $n=2k$, $\lambda\in \F_{2^k}^*$ and $(u,v)\in \F^*_{2^n}\times \F^*_{2^n}$.
 Then $f(x)=\Tr_1^k(\lambda x^{2^k+1})+\Tr_1^n(ux)\Tr_1^n(vx)$
is bent-negabent on $\F_{2^n}$ if and only if one of the following  conditions is
satisfied:
\begin{enumerate}
  \item $\lambda\neq 1,$ \, $(\Tr_1^n(\frac{u}{1+\lambda}),\Tr_1^n(\frac{(\lambda u^{2^k}+u)v}{1+\lambda^2}),
  \Tr_1^n(\lambda^{-1}u^{2^k}v))=(0,0,0)$ or
 $( \Tr_1^n(\frac{u}{1+\lambda}),$ $ \Tr_1^n(\frac{(\lambda u^{2^k}+u+1+\lambda)v}{1+\lambda^2}), $ $
  \Tr_1^n(\lambda^{-1}u^{2^k}v))=(1,0,0)$;
    \item $\lambda= 1, \, k=2,\, u,v,u+v\notin \F_{2^k}$
and $ \Tr_1^n(u^{2^k}v)=0$.
\end{enumerate}
\end{corollary}

As a special case of Theorem \ref{negapoly1}, if $\lambda=0$, then it gives the necessary and sufficient conditions for $\Tr_1^n(ux)\Tr_1^n(vx) $ to be negabent on $\F_{2^n}$ for even $n$.  In the following we consider the negabent property of $\Tr_1^n(ux)\Tr_1^n(vx) $ for both even and odd $n$.

\begin{theorem}\label{negapoly2}
Let $f(x)= \Tr_1^n(ux)\Tr_1^n(vx) $, where $(u,v)\in \F^*_{2^n}\times \F^*_{2^n}$. Then $f(x)$ is negabent on $\F_{2^n}$ if and only if one of the following conditions is satisfied:
\begin{enumerate}
\item $\Tr_1^n(u)=0$ and  $\Tr_1^n(uv)=0$;
\item $\Tr_1^n(u)=1$ and  $\Tr_1^n((u+1)v)=0$.
\end{enumerate}
\end{theorem}

{\em Proof:}
 According to Theorem \ref{negath}, it is sufficient to prove that
$$f(x)+f(x+a)+\Tr_1^n(ax)=\Tr_1^n\Big((\Tr_1^n(va)u+\Tr_1^n(ua)v+a)x \Big)+\Tr_1^n(ua)\Tr_1^n(va)$$
 is balanced
for all nonzero $a\in \F_{2^n}$,
 which is equivalent to show that  $ \Tr_1^n(va)u+\Tr_1^n(ua)v+a\neq 0$
 for all nonzero $a$. Let $h(a)=\Tr_1^n(va)u+\Tr_1^n(ua)v+a,$ we have
 \begin{enumerate}
  \item [1)] $(\Tr_1^n(ua), \Tr_1^n(va))=(0,0)$: For this case, $h(a)=0$ has the only solution $a=0$.
  \item [2)] $(\Tr_1^n(ua), \Tr_1^n(va))=(0,1)$: In this case, $ h(a)= 0$  has the only
  solution $a=u$ if and only if $\Tr_1^n(u)=0$ and $\Tr_1^n(uv)=1 $.
  \item [3)] $(\Tr_1^n(ua), \Tr_1^n(va))=(1,0)$: Similar as above, for this case $h(a)=0$ has
  the only solution $a=v$ if and only if $\Tr_1^n(uv)=1 $ and $\Tr_1^n(v)=0$.
  \item [4)] $(\Tr_1^n(ua), \Tr_1^n(va))=(1,1)$: In this case, $a=u+v$
  is the only solution to
 $ \Tr_1^n(va)u+\Tr_1^n(ua)v+a= 0$ if and only if $\Tr_1^n(u(u+v))=1 $ and $\Tr_1^n(v(u+v))=1$.
\end{enumerate}
%
Based on Cases 1)-4), it can be seen that $ \Tr_1^n(va)u+\Tr_1^n(ua)v+a\neq 0$
 for all nonzero $a$ if and only if
 one  of the two conditions in Theorem \ref{negapoly2} is satisfied.
\done

\begin{remark}
Theorem \ref{negapoly2} shows that  $\Tr_1^n(x)\Tr_1^n(vx) $ is negabent for any nonzero $v\in\F_{2^n}$ when $n$ is odd and $u=1$,
which was given in Theorem 8 in \cite{zhou15}. Note that the negabent property is not preserved by linear transform, i.e.,
$f(x)$ is negabent on $\F_{2^n}$ does not imply that $f(ax)$ is negabent on $\F_{2^n}$ for all $a\in \F_{2^n}^*$ \cite{matthew08}. Thus,
Theorem \ref{negapoly2} is  not a special case  of Theorem 8 in \cite{zhou15}.
\end{remark}
%
%

\begin{theorem}\label{negapoly3}
Let $n$ be an even integer and $k$ be a positive integer such that $\gcd(n,3k)=\gcd(n,k)$.
 Then $f(x)=\Tr_1^n( x^{2^k+1})+\Tr_1^n(x)\Tr_1^n(vx)$
is negabent on $\F_{2^n}$ if $\Tr_1^n(v)=0$.
\end{theorem}
{\em Proof:}
According to Theorem \ref{negath},
we only need to show that  $f(x)+f(x+a)+\Tr_1^n(ax)$ is balanced for all nonzero $a\in\F_{2^n}$ if $\Tr_1^n(v)=0$.
A direct calculation gives
 \begin{eqnarray*}
 f(x)+f(x+a)+\Tr_1^n(ax) &=&\Tr_1^n(a^{2^k}x+ax^{2^k})+\Tr_1^n(a)\Tr_1^n(vx)+\Tr_1^n(va)\Tr_1^n(x)+\Tr_1^n(ax)\\
 &&+\Tr_1^n( a^{2^k+1})
 +\Tr_1^n(a)\Tr_1^n(va)\\
 &=&\Tr_1^n((a^{2^k}+  a^{2^{-k}} +a)x)+\Tr_1^n((v\Tr_1^n(a))x)+\Tr_1^n(\Tr_1^n(va)x)\\
 &&+\Tr_1^n(a^{2^k+1})
 +\Tr_1^n(a)\Tr_1^n(va).
 \end{eqnarray*}
 This shows that $f(x)+f(x+a)+\Tr_1^n(ax)$ is balanced if and only if $ a^{2^k}+  a^{2^{-k}} +a+v\Tr_1^n(a) +\Tr_1^n(va)\neq 0$,
 i.e., $ a+ a^{2^k} +a^{2^{2k}}+v^{2^k}\Tr_1^n(a) +\Tr_1^n(va)\neq 0$. Notice that $a+ a^{2^k} +a^{2^{2k}}$ is a permutation of $\F_{2^n}$ due to Lemma \ref{permu1}.
 Let $g(a)=a+ a^{2^k} +a^{2^{2k}}+v^{2^k}\Tr_1^n(a) +\Tr_1^n(va)$ and $h(a)$ be the compositional inverse of $a+ a^{2^k} +a^{2^{2k}}$, then we have
 \begin{enumerate}
  \item [1)] $(\Tr_1^n(a), \Tr_1^n(va))=(0,0)$: For this case, $g(a)=0$ has the only solution $a=0$.
  \item [2)] $(\Tr_1^n(a), \Tr_1^n(va))=(0,1)$: In this case, $ g(a)= 0$  means that  $a+ a^{2^k} +a^{2^{2k}}=1$,
  i.e., $a=h(1)=1$. However, $ \Tr_1^n(va)=\Tr_1^n(v)=0$, which shows that $ g(a)= 0$  has no solution in this case.
  \item [3)] $(\Tr_1^n(a), \Tr_1^n(va))=(1,0)$:
  In this case, $ g(a)= 0$ is reduced to $a+ a^{2^k} +a^{2^{2k}}=v^{2^k}$,
  i.e., $a=h(v^{2^k})$. However, by Lemma \ref{permu1},  $ \Tr_1^n(a)=\Tr_1^n(h(v^{2^k}))=\Tr_1^n(v^{2^k})=\Tr_1^n(v)=0$.
 This shows that $ g(a)= 0$  has no solution in this case.
  \item [4)] $(\Tr_1^n(a), \Tr_1^n(va))=(1,1)$: Similar as above,
  $ g(a)= 0$  implies   that  $a+ a^{2^k} +a^{2^{2k}}=1+v^{2^k}$,
  i.e., $a=h(1+v^{2^k})$. Note that $\Tr_1^n(1)=0$ since $n$ is even. From Lemma \ref{permu1}, $ \Tr_1^n(a)=\Tr_1^n(h(1+v^{2^k}))=\Tr_1^n(1+v^{2^k})=\Tr_1^n(v)=0$,
  which shows that $ g(a)= 0$  has no solution in this case.
\end{enumerate}
From the above Cases 1)-4), we can see that $ a+ a^{2^k} +a^{2^{2k}}+v^{2^k}\Tr_1^n(a) +\Tr_1^n(va)\neq 0$ for all nonzero $a\in\F_{2^n}$ if $\Tr_1^n(v)=0$.
This completes the proof. \done

By the same techniques used in the proof of Theorem \ref{negapoly3}, we can derive the following result.

\begin{theorem}\label{negapoly4}
Let $r$ and  $k$ be two  integers such that $rk$ is even.
Let $n=rk$,  $\lambda\in \F^*_{2^k}$ and $\gcd(\lambda+x+\lambda x^{2},x^r-1)=1$.
 Then $f(x)=\Tr_1^n(\lambda x^{2^k+1})+\Tr_1^n(x)\Tr_1^n(vx)$
is negabent on $\F_{2^n}$ if $\Tr_1^n(v)=0$.
\end{theorem}
{\em Proof:}
According to Theorem \ref{negath},
it is enough to prove   that  $f(x)+f(x+a)+\Tr_1^n(ax)$ is balanced for all nonzero $a\in\F_{2^n}$ for the $v\in\F_{2^n}$ satisfying $\Tr_1^n(v)=0$.
Note that
 \begin{eqnarray*}
 f(x)+f(x+a)+\Tr_1^n(ax) &=&\Tr_1^n(\lambda(a^{2^k}x+ax^{2^k}))+\Tr_1^n(a)\Tr_1^n(vx)+\Tr_1^n(va)\Tr_1^n(x)+\Tr_1^n(ax)\\
 &&+\Tr_1^n(\lambda a^{2^k+1})
 +\Tr_1^n(a)\Tr_1^n(va)\\
 &=&\Tr_1^n((\lambda a^{2^k}+ (\lambda a)^{2^{-k}} +a)x)+\Tr_1^n(v\Tr_1^n(a)x)+\Tr_1^n(\Tr_1^n(va)x)\\
 &&+\Tr_1^n(\lambda a^{2^k+1})
 +\Tr_1^n(a)\Tr_1^n(va).
 \end{eqnarray*}
 Thus, $f(x)+f(x+a)+\Tr_1^n(ax)$ is balanced if and only if
  \begin{eqnarray}\label{eqth6}
   \lambda a^{2^k}+ (\lambda a)^{2^{-k}} +a +v\Tr_1^n(a)+\Tr_1^n(va) \neq 0.
   \end{eqnarray}
   Raising both sides of \eqref{eqth6} to the $2^k$-th power, we get
  $\lambda a^{2^{2k}}+ \lambda a+a^{2^k}+ v^{2^k}\Tr_1^n(a)+\Tr_1^n(va) \neq 0$ due to $\lambda\in\F_{2^k}$.
 Let $g(a)=\lambda a+a^{2^k}+\lambda a^{2^{2k}}  +v^{2^k}\Tr_1^n(a)+\Tr_1^n(va)$.
According to Lemma \ref{permu2},   $\lambda a+a^{2^k}+\lambda a^{2^{2k}} $ is a permutation of
$\F_{2^n}$ since  $\gcd(\lambda+x+\lambda x^{2},x^r-1)=1$.
Let $h(a)$ be the compositional inverse of $\lambda a+a^{2^k}+\lambda a^{2^{2k}} $.
Similar as in the proof of Theorem \ref{negapoly3},
we have
 \begin{enumerate}
  \item [1)] $(\Tr_1^n(a), \Tr_1^n(va))=(0,0)$: For this case, $g(a)=0$ has the only solution $a=0$.
  \item [2)] $(\Tr_1^n(a), \Tr_1^n(va))=(0,1)$: In this case, $ g(a)= 0$  means that  $\lambda a+a^{2^k}+\lambda a^{2^{2k}} =1$,
  i.e., $a=h(1)=1$ since $\lambda \cdot 1+1^{2^k}+\lambda \cdot 1^{2^{2k}}=1$. However, $ \Tr_1^n(va)=\Tr_1^n(v)=0$, which shows that $ g(a)= 0$  has no solution in this case.
  \item [3)] $(\Tr_1^n(a), \Tr_1^n(va))=(1,0)$:
  In this case, $ g(a)= 0$  means   that  $\lambda a+a^{2^k}+\lambda a^{2^{2k}}=v^{2^k}$,
  i.e., $a=h(v^{2^k})$. From Lemma \ref{permu2},  $ \Tr_1^n(a)=\Tr_1^n(h(v^{2^k}))=\Tr_1^n(v^{2^k})=\Tr_1^n(v)=0$,
  which shows that $ g(a)= 0$  has no solution in this case.
  \item [4)] $(\Tr_1^n(a), \Tr_1^n(va))=(1,1)$: Similar as above,
  $ g(a)= 0$  implies   that  $\lambda a+a^{2^k}+\lambda a^{2^{2k}}=1+v^{2^k}$,
  i.e., $a=h(1+v^{2^k})$. Note that $\Tr_1^n(1)=0$ due to $n$ is even. Again by Lemma \ref{permu2}, $ \Tr_1^n(a)=\Tr_1^n(h(1+v^{2^k}))=\Tr_1^n(1+v^{2^k})=\Tr_1^n(v)=0$.
 This implies that $ g(a)= 0$  has no solution in this case.
\end{enumerate}
From the above Cases 1)-4), we can see that if $\Tr_1^n(v)=0$, then
$ \lambda a+a^{2^k}+\lambda a^{2^{2k}}+v^{2^k}\Tr_1^n(a) +\Tr_1^n(va)\neq 0$ for all nonzero $a\in\F_{2^n}$.
This completes the proof. \done

\begin{remark}
Notice that if one takes $n=rk$ in Theorem \ref{negapoly3} then Theorem \ref{negapoly3} is a special case of Theorem \ref{negapoly4} due to the fact that $\gcd(1+x+x^2,x^r-1)=1$ if and only if $\gcd(rk,3k)=\gcd(rk,k)$. For the values of $n, k$ with $\gcd(n,k)\not=k$, the results in Theorem \ref{negapoly3} are not covered by Theorem \ref{negapoly4}.
\end{remark}

By Theorem \ref{negapoly4} we can obtain the following results if we take $r=3, 4, 5$ respectively.

\begin{corollary}
Let $k$ be an even integer and $n=3k$.
 Let $\lambda\in\F_{2^k} \setminus \{0, 1\}$.
 Then $f(x)=\Tr_1^n(\lambda x^{2^k+1})+\Tr_1^n(x)\Tr_1^n(vx)$
is negabent on $\F_{2^n}$ if $\Tr_1^n(v)=0$.
\end{corollary}
{\em Proof:}
According to Theorem \ref{negapoly4}, it is sufficient to show that $\gcd(\lambda+x+\lambda x^{2}, x^3-1 )=1$ if
 $\lambda\ne 1$. Then result follows from the fact that $\gcd(\lambda+x+\lambda x^{2}, x^3-1 )=\gcd(\lambda+x+\lambda x^{2}, x^2+x+1 )=\gcd(\lambda(x^2+x+1)+(\lambda+1)x, x^2+x+1 )=\gcd((\lambda+1)x, x^2+x+1)$.
 \done

If $r=4$, then $\gcd(\lambda+x+\lambda x^{2}, x^4-1)=\gcd(\lambda+x+\lambda x^{2}, x-1)=1$ for any $\lambda\in\F_{2^k}^*$. Thus, we have
\begin{corollary}
Let  $n=4k$ and  $\lambda\in\F_{2^k}^* $.
 Then $f(x)=\Tr_1^n(\lambda x^{2^k+1})+\Tr_1^n(x)\Tr_1^n(vx)$
is negabent on $\F_{2^n}$ if $\Tr_1^n(v)=0$.
\end{corollary}

\begin{corollary} Let $k$ be an even integer and $n=5k$.
Let $\lambda\in\F_{2^k} \setminus \{0, \omega,\omega^2\}$, where $\omega$ is
a primitive element of  $\F_{2^2}.$
 Then $f(x)=\Tr_1^n(\lambda x^{2^k+1})+\Tr_1^n(x)\Tr_1^n(vx)$
is negabent on $\F_{2^n}$ if $\Tr_1^n(v)=0$.
\end{corollary}

{\em Proof:}
According to Theorem \ref{negapoly4}, we need to determine the condition on $\lambda$ such that $\gcd(\lambda+x+\lambda x^{2}, x^5-1 )=1$. Notice that $\gcd(\lambda+x+\lambda x^{2}, x^5-1 )=\gcd(\lambda+x+\lambda x^{2}, x^4+x^3+x^2+x+1)$. By a simple calculation, we have
$x^4+x^3+x^2+x+1=(1+\mu x+ x^2)(\mu^2+\mu+(\mu+1)x+x^2)+(\mu^2+\mu+1)(\mu x+1)$,
where $\mu=\lambda^{-1}$. This leads to $\gcd(\lambda+x+\lambda x^{2}, x^4+x^3+x^2+x+1)=\gcd(1+\mu x+ x^2, x^4+x^3+x^2+x+1)=\gcd(1+\mu x+ x^2,(\mu^2+\mu+1)(\mu x+1))=1$ if and only of $\mu^2+\mu+1\ne 0$. This completes the proof.
\done

\section{On a class of monomial negabent functions}\label{secnegamono}
In  \cite{zhou15}, Zhou and Qu showed that
$\Tr_1^{2k}(\lambda x^d)$ is negabent on $\F_{2^{2k}}$ if $\lambda\in \F_2,$ where
 $d=2^{k}+3$ and $k\geq 3$ is odd.
In this section, we will show that  $\lambda\in \F_2$ is also necessary for
$\Tr_1^{2k}(\lambda x^d)$ to be  negabent.



%

%
%

%

\begin{theorem}\label{qplus3}
Let $n=2k$, $q=2^k $ and $d=q+3$, where  $k\geq 3$ is  odd.
 Then $\Tr_1^n(\lambda x^d)$ is  negabent on $\F_{2^n}$ if and only if $\lambda\in \F_2$.
\end{theorem}

{\em Proof:}
Since $k$ is odd,  then $f(x)=x^2+x+1$ is irreducible over $\F_{2^k}$ as it is irreducible over $\F_2$.
Let $\omega$ be a root of $f(x)$. Then ${\mathbb F}_{2^n}={\mathbb F}_{2^{k}}[\omega],$ i.e., each $x\in{\mathbb F}_{2^n}$ can be uniquely represented as
$
x_0+x_1\omega,
$
where $x_i\in {\mathbb F}_{2^{k}}.$
  Then
\begin{eqnarray}\label{nq+3fx}
 x^d=(x_0+x_1\omega)^d=x_0^4+x_1^4+x_1x_0^3+x_0x_1^3+(x_0^2x_1^2+x_0x_1^3+x_1^4)\omega
\end{eqnarray}   and
\begin{eqnarray}\label{nq+3fx+a}
 (x+a)^d&=&(x_0+a_0)^4+(x_1+a_1)^4+(x_1+a_1)(x_0+a_0)^3+(x_0+a_0)(x_1+a_1)^3\nonumber\\
&&+((x_0+a_0)^2(x_1+a_1)^2+(x_0+a_0)(x_1+a_1)^3+(x_1+a_1)^4)\omega,
\end{eqnarray} where $a=a_0+a_1\omega$.

Note that $ \Tr_{k}^{2k}(1)=0$ and $ \Tr_{k}^{2k}(\omega)=\omega+\omega^{2^k}=1$ since $k$ is odd and $\omega$ is a root of $x^2+x+1$.
Let  $\lambda=\lambda_0+\lambda_1\omega.$  Then from \eqref{nq+3fx}, \eqref{nq+3fx+a} and $\Tr_k^{2k}(ax)=a_0x_1+a_1x_0+a_1x_1,$  we have
\begin{eqnarray}\label{nqtr}
&& \Tr_k^{2k}(\lambda x^d+\lambda (x+a)^d+ax)\nonumber\\
&=&\lambda_1x_0^2a_1^2+\lambda_0x_0a_1^3+\lambda_0x_0^2a_1^2+\lambda_1a_1x_0^3+\lambda_0a_0x_1^3
+a_1x_0+\lambda_1x_1x_0a_0^2+\lambda_1x_1x_0^2a_0+\lambda_1a_1x_0a_0^2\nonumber\\
&&+\lambda_1a_1x_0^2a_0+x_1a_1+\lambda_0a_0x_1^2a_1+\lambda_0x_0x_1^2a_1+\lambda_0x_0x_1a_1^2+\lambda_0a_0x_1a_1^2+\lambda_0a_0a_1^3+\lambda_0a_1^4\nonumber\\
&&+\lambda_1a_0^4+\lambda_1x_1a_0^3+\lambda_1a_1a_0^3+\lambda_1a_0^2x_1^2+\lambda_1a_0^2a_1^2+\lambda_0a_0^2x_1^2+\lambda_0a_0^2a_1^2+a_0x_1=G(x_0,x_1).
\end{eqnarray}

Suppose that $\lambda_1\neq 0. $  We will show that for each $\lambda=\lambda_0+\lambda_1\omega$
with $\lambda_1\neq0$, there exists at least one nonzero $a=a_0+a_1\omega\in\F_{2^n}$ such that
$\Tr_1^n(\lambda x^d+\lambda (x+a)^d+ax)=\Tr_1^k(G(x_0,x_1))$ is not balanced. We consider this  in three cases.

Case (i)   $ \lambda_1\neq0, \lambda_0^2+\lambda_1^2+\lambda_0\lambda_1+1\neq 0$.

In this case, let $a_1=0$ and $a_0\neq 0$. Then
\begin{eqnarray}\label{casea}
&&\sum_{x_0,x_1\in\F_q}(-1)^{\Tr_1^k(G(x_0,x_1))}\nonumber\\
&=& \sum_{x_0,x_1\in\F_q}(-1)^{\Tr_1^k(\lambda_1x_1x_0^2a_0+\lambda_1x_1x_0a_0^2+\lambda_1a_0^2x_1^2+\lambda_0a_0x_1^3+\lambda_0a_0^2x_1^2+\lambda_1a_0^4+\lambda_1x_1a_0^3+a_0x_1)}\nonumber\\
&=& \sum_{x_1\in\F_q}(-1)^ {\Tr_1^k(\lambda_1a_0^2x_1^2+\lambda_0a_0x_1^3+\lambda_0a_0^2x_1^2+\lambda_1a_0^4+\lambda_1x_1a_0^3+a_0x_1)}
\sum_{x_0\in\F_q}(-1)^{\Tr_1^k((\lambda_1x_1a_0+\lambda_1^2x_1^2a_0^4)x_0^2) }\nonumber\\
&=&2^k\sum_{x_1=0 \,\texttt{or}\,  x_1=(\lambda_1a_0^3)^{-1} }(-1)^{\Tr_1^k(\lambda_1a_0^2x_1^2+\lambda_0a_0x_1^3+\lambda_0a_0^2x_1^2+\lambda_1a_0^4+\lambda_1x_1a_0^3+a_0x_1)}\nonumber\\
&=&2^k\big((-1)^{\Tr_1^k(\lambda_1a_0^4)}+(-1)^{\Tr_1^k(\lambda_1a_0^2t^2+\lambda_0a_0t^3+\lambda_0a_0^2t^2+\lambda_1a_0^4+\lambda_1ta_0^3+a_0t)}\big),
\end{eqnarray} where $t=(\lambda_1a_0^3)^{-1}.$
 By \eqref{casea}, if there exists $a_0\in \F^*_{q}$ such that
 $\Tr_1^k(\lambda_1a_0^2t^2+\lambda_0a_0t^3+\lambda_0a_0^2t^2+\lambda_1ta_0^3+a_0t)=0,$  then
 $\sum_{x_0,x_1\in\F_q}(-1)^{\Tr_1^k(G(x_0,x_1))}=(-1)^{\Tr_1^k(\lambda_1a_0^4) }\cdot 2^{k+1}\neq0,$ i.e., $\Tr_1^k(G(x_0,x_1))$ is not balanced for such $a_0\in \F^*_{q}$.
 Since $t=(\lambda_1a_0^3)^{-1}$, we have
  $\Tr_1^k(\lambda_1a_0^2t^2+\lambda_0a_0t^3+\lambda_0a_0^2t^2+\lambda_1ta_0^3+a_0t)=
 \Tr_1^k(\frac{\lambda_1^2+\lambda_0^2+1+\lambda_0\lambda_1}{\lambda_1^4}(a_0^8)^{-1}+1)$,
  which implies that there exists $a_0\neq0$ such that $\Tr_1^k(\frac{\lambda_1^2+\lambda_0^2+1+\lambda_0\lambda_1}{\lambda_1^4}(a_0^8)^{-1})+1=0$ if $\lambda\in \F_{2^n}$ satisfying $ \lambda_0^2+\lambda_1^2+\lambda_0\lambda_1+1\neq 0$ and $\lambda_1\neq 0$.

Case (ii)    $\lambda_1\neq0,  \lambda_0^2+\lambda_1^2+\lambda_0\lambda_1+1= 0$ and $\lambda_0\neq0$.

In this case, let $a_0=0$ and $a_1\neq 0.$
 Then
\begin{eqnarray}\label{caseb}
&&\sum_{x_0,x_1\in\F_q}(-1)^{\Tr_1^k(G(x_0,x_1))}\nonumber\\
&=& \sum_{x_0,x_1\in\F_q}(-1)^{\Tr_1^k(\lambda_0x_0x_1^2a_1+(\lambda_0a_1^2x_0+a_1)x_1+\lambda_1a_1x_0^3+(\lambda_1a_1^2+\lambda_0a_1^2)x_0^2+(\lambda_0a_1^3+a_1)x_0+\lambda_0a_1^4)}\nonumber\\
&=& \sum_{x_0\in\F_q}(-1)^ {\Tr_1^k(\lambda_1a_1x_0^3+(\lambda_1a_1^2+\lambda_0a_1^2)x_0^2+(\lambda_0a_1^3+a_1)x_0+\lambda_0a_1^4)}
\sum_{x_1\in\F_q}(-1)^{\Tr_1^k((\lambda_0x_0a_1+\lambda_0^2a_1^4x_0^2+a_1^2)x_1^2) }\nonumber\\
&=&2^k\sum_{x_0=y_1 \,\texttt{or}\,  x_0=y_2 }(-1)^{\Tr_1^k(\lambda_1a_1x_0^3+(\lambda_1a_1^2+\lambda_0a_1^2)x_0^2+(\lambda_0a_1^3+a_1)x_0+\lambda_0a_1^4)},
\end{eqnarray} where $y_1$ and $y_2$ are the two roots of $\lambda_0x_0a_1+\lambda_0^2a_1^4x_0^2+a_1^2=0$
($x_0$ as the indeterminate variable)
under the condition $\Tr_1^k(a_1)=0.$
Thus, $y_1+y_2=\frac{1}{\lambda_0a_1^3}$ and $y_1y_2=\frac{1}{\lambda_0^2a_1^2}$.
 By \eqref{caseb}, if there exists $a_1\in \F^*_{q}$ such that $\Tr_1^k(a_1)=0$ and
 $\Tr_1^k(\lambda_1a_1(y_1^3+y_2^3)+(\lambda_1a_1^2+\lambda_0a_1^2)(y_1+y_2)^2+(\lambda_0a_1^3+a_1)(y_1+y_2))=0,$
   then
 $\sum_{x_0,x_1\in\F_q}(-1)^{\Tr_1^k(G(x_0,x_1))}=\pm 2^{k+1}\neq0,$ i.e., $\Tr_1^k(G(x_0,x_1))$ is not balanced for such $a_1\in \F^*_{q}$.
 By $ y_1^3+y_2^3=(y_1+y_2)^3+y_1y_2(y_1+y_2)=\frac{1}{\lambda_0^3}(\frac{1}{a_1^9}+\frac{1}{a_1^5})$, one obtains that
  \begin{eqnarray*}
 && \Tr_1^k(\lambda_1a_1(y_1^3+y_2^3)+(\lambda_1a_1^2+\lambda_0a_1^2)(y_1+y_2)^2+(\lambda_0a_1^3+a_1)(y_1+y_2))\\
&=&\Tr_1^k((\frac{\lambda_1^2}{\lambda_0^6}+\frac{\lambda_0^2+\lambda_1^2+\lambda_0\lambda_1+1}{\lambda_0^4})\frac{1}{a_1^8}+1)= \Tr_1^k(\frac{\lambda_1^2}{\lambda_0^6}\cdot\frac{1}{a_1^8}+1).
 \end{eqnarray*}
 According to Lemma \ref{lemma-kloo}, for odd $k>2$, there exists $a_1\in\F_q^*$ such that $\Tr_1^k(\frac{\lambda_1^2}{\lambda_0^6}\cdot\frac{1}{a_1^8}+1)=\Tr_1^k((\frac{\lambda_1^2}{\lambda_0^6})^{-8}\cdot\frac{1}{a_1})+1=0$
 and  $\Tr_1^k(a_1)=0$.
 Thus, for any
 $\lambda\in \F_{2^n}$ such that
 $ \lambda_0^2+\lambda_1^2+\lambda_0\lambda_1+1= 0$ and $\lambda_0\lambda_1\neq 0 $, there exists $a_1\neq0$ such that $\Tr_1^k(a_1)=0$ and
 $\Tr_1^k(\lambda_1a_1(y_1^3+y_2^3)+(\lambda_1a_1^2+\lambda_0a_1^2)(y_1+y_2)^2+(\lambda_0a_1^3+a_1)(y_1+y_2))=0.$ That is, $\Tr_1^k(G(x_0,x_1))$ is not balanced for such $a_1\in \F^*_{q}$.

Case (iii)    $\lambda_1\neq0,  \lambda_0^2+\lambda_1^2+\lambda_0\lambda_1+1= 0$ and $\lambda_0=0$.

For this case, $\lambda_1=1$ and $\lambda_0=0$.
Let $a_0=a_1\neq 0.$ Then
\begin{eqnarray}\label{casec}
&&\sum_{x_0,x_1\in\F_q}(-1)^{\Tr_1^k(G(x_0,x_1))}\nonumber\\
&=& \sum_{x_0,x_1\in\F_q}(-1)^{\Tr_1^k(a_0^2x_1^2+(x_0^2a_0+x_0a_0^2+a_0^3)x_1+x_0^3a_0+(a_0^3+a_0)x_0+a_0^4)}\nonumber\\
&=& \sum_{x_0\in\F_q}(-1)^ {\Tr_1^k(x_0^3a_0+(a_0^3+a_0)x_0+a_0^4)}
\sum_{x_1\in\F_q}(-1)^{\Tr_1^k((a_0+x_0^2a_0+x_0a_0^2+a_0^3)x_1) }\nonumber\\
&=&2^k\sum_{x_0=y_1 \,\texttt{or}\,  x_0=y_2 }(-1)^{\Tr_1^k(x_0^3a_0+(a_0^3+a_0)x_0+a_0^4)},
\end{eqnarray} where $y_1$ and $y_2$ are the two roots of $a_0+x_0^2a_0+x_0a_0^2+a_0^3=0$
($x_0$ as the indeterminate variable)
 under the condition $\Tr_1^k(a_0^{-1})=1.$
Thus, $y_1+y_2=a_0$ and $y_1y_2=1+a_0^2$.
 By \eqref{casec}, if there exists $a_0\in \F^*_{q}$ such that $\Tr_1^k(a_0^{-1})=1$ and
 $ \Tr_1^k((y_1^3+y_2^3)a_0+(a_0^3+a_0)(y_1+y_2))=0, $
   then
 $\sum_{x_0,x_1\in\F_q}(-1)^{\Tr_1^k(G(x_0,x_1))}=\pm 2^{k+1}\neq0.$ That is, $\Tr_1^k(G(x_0,x_1))$ is not balanced for such $a_0\in \F^*_{q}$.
  Note that $ y_1^3+y_2^3=(y_1+y_2)^3+y_1y_2(y_1+y_2)=a_0^3+(1+a_0^2)a_0=a_0$, then
  $\Tr_1^k((y_1^3+y_2^3)a_0+(a_0^3+a_0)(y_1+y_2))=\Tr_1^k(a_0^2+(a_0^3+a_0)a_0)=\Tr_1^k(a_0)=0 $. Again by
  Lemma \ref{lemma-kloo}, for odd $k>2$, there exists $a_0\in\F_q^*$ such that $\Tr_1^k(a_0)=0$ and $\Tr_1^k(a_0^{-1})=1$.
 Thus, for
 $\lambda=\lambda_0+\lambda_1\omega=\omega$, there exists $a_0\neq0$ such that $\Tr_1^k(a_0^{-1})=1$ and $\Tr_1^k((y_1^3+y_2^3)a_0+(a_0^3+a_0)(y_1+y_2))=0$,  which implies that $\Tr_1^k(G(x_0,x_1))$ is not balanced.

From the above Cases (i)-(iii), for each $\lambda=\lambda_0+\lambda_1\omega$
with $\lambda_1\neq0$, there exists at least one nonzero $a=a_0+a_1\omega\in\F_{2^n}$ such that
$\Tr_1^n(\lambda x^d+\lambda (x+a)^d+ax)=\Tr_1^k(G(x_0,x_1))$ is not balanced.

 In the following we assume that $\lambda_1=0 $
and $\lambda=\lambda_0+\lambda_1\omega=\lambda_0\neq0. $ Let $a_1=0$.
Then
\begin{eqnarray}\label{cased1}
 \sum_{x_0,x_1\in\F_q}(-1)^{\Tr_1^k(G(x_0,x_1))}\nonumber
&=& \sum_{x_0,x_1\in\F_q}(-1)^{\Tr_1^k(\lambda_0a_0x_1^3+\lambda_0a_0^2x_1^2+a_0x_1)}\nonumber\\
&=& 2^k\sum_{x_1\in\F_q}(-1)^ {\Tr_1^k(\lambda_0a_0x_1^3+\lambda_0a_0^2x_1^2+a_0x_1)}\nonumber\\
&=& 2^k\sum_{x_1\in\F_q}(-1)^ {\Tr_1^k(\lambda_0a_0x_1^3+(\lambda_0^{2^{k-1}}a_0+a_0)x_1)}.
\end{eqnarray}
Since $k$ is odd, then $\gcd(3,2^k-1)=1$. Let $ \lambda_0=r^3, a_0=t^3 $, then from \eqref{cased1}, one gets
\begin{eqnarray}\label{cased2}
\sum_{x_0,x_1\in\F_q}(-1)^{\Tr_1^k(G(x_0,x_1))}
= 2^k\sum_{x_1\in\F_q}(-1)^ {\Tr_1^k(x_1^3+(r^{3\cdot 2^{k-1}}+1)r^{-1}t^2x_1)}.
\end{eqnarray}
Thus, if $\lambda_0=r^3\neq 1,$ then $r^{3\cdot 2^{k-1}}+1\neq0$.
We claim that for any $r\in\F_q^* $ and $r\neq 1$,  there must exist some
$a_0\in\F_q^*$ such that
$\sum_{x_0,x_1\in\F_q}(-1)^{\Tr_1^k(G(x_0,x_1))}=
2^k\sum_{x_1\in\F_q}(-1)^ {\Tr_1^k(x_1^3+(r^{3\cdot 2^{k-1}}+1)r^{-1}t^2x_1)}\neq 0$, i.e., $\Tr_1^k(G(x_0,x_1))$ is not balanced.
 Otherwise, the Walsh-Hadamard  transform of $\Tr_1^k(x^3)$ at any point $t\in \F_q$ is zero,
  which contradicts with Parseval's theorem\footnote{Parseval's theorem shows that for
any   Boolean function $f(x)$ from $\F_{2^k}$ to $\F_2$,  its Walsh-Hadamard transform $W_f(u)$
  satisfies $\sum_{u\in\F_{2^k}}(W_f(u))^2=2^{2k}$.}.

Therefore, if $\Tr_1^n(\lambda x^d)$ is negabent on $\F_{2^n}$, then $\lambda$ has to be in $\F_2$. Zhou and Qu \cite[Theorem 6]{zhou15} proved  that if $\lambda\in\F_2$, then $\Tr_1^n(\lambda x^d)$ is indeed negabent on $\F_{2^n}.$  This completes the proof.
\done

To end this section, we present a conjecture on
negabent monomials whose exponents are of Niho type, namely the exponents of the form $d=r(2^m-1)+1$,
where $m=n/2$ and $1\leq r\leq 2^m$. Notice that $d_1=r_1(2^m-1)+1$ and $d_2=r_2(2^m-1)+1$ lie in the same cyclotomic coset modulo $2^n-1$ if and only if
$r_1\equiv r_2\pmod{2^m+1}$ or $r_1+r_2\equiv 1\pmod{2^m+1}$.

 Sarkar \cite{sarkar12} gave a class of negabent monomials whose  exponents are of Niho type, as follows:

\begin{theorem}\label{qplus1}\cite{sarkar12}
Let $n=2m$ and $d=(2^{m-1}+1)(2^m-1)+1$.
Then $\Tr_1^n(\alpha x^d)$ is negabent if and only if
$\alpha+\alpha^{2^m}\neq 1$.
\end{theorem}

Based on  our computer experiments,  we have the following conjecture:

\begin{conjecture}\label{conj1}
Let $n=2m$ and $d=r(2^m-1)+1$, where $2 \leq r \leq 2^{m-1}+1$.
   Then $\Tr_1^n(\alpha x^d)$ is a negabent function if and only if
one of the following two  conditions holds:
\begin{enumerate}
  \item $m$ is odd,  $r=2^{m-2}+1\equiv \frac{3}{4}\pmod {2^m+1}$ and $\alpha\in \F_2$. (Cubic functions, Theorem \ref{qplus3})
  \item  $r=2^{m-1}+1\equiv \frac{1}{2}\pmod {2^m+1}$ and $\alpha+\alpha^{2^m}\neq 1$. (Quadratic functions, Theorem \ref{qplus1})
\end{enumerate}
\end{conjecture}
This conjecture has been verified by Magma for $ n \leq 14$.

%
%


\section{Conclusion}\label{seccon}
Negabent functions as a generalization of bent functions are very useful in
cryptography and coding theory. In this paper,
several classes of negabent functions of the form
  $f(x)=\Tr_1^n(\lambda x^{2^k+1})+\Tr_1^n(ux)\Tr_1^n(vx)$ were given, where $0<k<n$ and
   $(u,v)\in \F^*_{2^n}\times \F^*_{2^n}$. In particular,
   we gave
   the necessary and sufficient conditions
   for $\Tr_1^{k}(\lambda x^{2^k+1})+\Tr_1^{2k}(ux)\Tr_1^{2k}(vx)$
   to be negabent on $\F_{2^{2k}}$,
   where $\lambda\in \F_{2^k}$.
We also  showed that
the condition $\lambda\in \F_2$ for $\Tr_1^{2k}(\lambda x^{2^k+3})$ to be negabent
is necessary, where $k\ge 3$ is odd.
  Finally, based on our Magma results, we presented
  a conjecture on monomial negabent functions whose exponents are of Niho type.

%

\end{document}